# Integrated Artificial Neurons from Metal Halide Perovskites

Jeroen J. de Boer[1], Bruno Ehrler[1,*]

[1]Center for Nanophotonics, AMOLF, 1098 XG, Amsterdam, the Netherlands
[*]Correspondence: b.ehrler@amolf.nl

## Abstract

Hardware neural networks could perform certain computational tasks orders of magnitude more energy-efficiently than conventional computers. Artificial neurons are a key component of these networks and are currently implemented with electronic circuits based on capacitors and transistors. However, artificial neurons based on memristive devices are a promising alternative, owing to their potentially smaller size and inherent stochasticity. But despite their promise, demonstrations of memristive artificial neurons have so far been limited. Here we demonstrate a fully on-chip artificial neuron based on microscale electrodes and halide perovskite semiconductors as the active layer. By connecting a halide perovskite memristive device in series with a capacitor, the device demonstrates stochastic leaky integrate-and-fire behavior, with an energy consumption of 20 to 60 pJ per spike, lower than that of a biological neuron. We simulate populations of our neuron and show that the stochastic firing allows the detection of sub-threshold inputs. The neuron can easily be integrated with previously-demonstrated halide perovskite artificial synapses in energy-efficient neural networks.

## Introduction

Artificial intelligence-based systems have seen a rapid increase in their capabilities in a wide range of tasks, such as natural language processing,[1] image recognition,[2,3] and strategizing.[4,5] The increase in the performance of these systems is accompanied by an exponential increase in the computational power, and thus the energy consumption.[6] Neuromorphic computing addresses this issue by implementing neural networks in hardware, lowering the required energy by orders of magnitude compared to conventional computers.[7] Neuromorphic chips rely on two main components for computation: artificial neurons, which integrate incoming signals and fire a voltage pulse upon reaching a threshold, and artificial synapses, which determine the connection strength between neurons. Ideally, both components can be integrated into a single chip in a dense arrangement to enable large-scale artificial neural networks. Both the neurons and synapses are typically implemented with electronic circuits composed of transistors and capacitors.[8] On the other hand, implementations that use memristive elements, which change their resistance based on an applied voltage, can be more compact and highly energy efficient, making them an attractive alternative.[9] Much research has gone into developing artificial synapses that directly use the resistance change of a memristive element as a proxy for connection strength.[9–12] Memristive elements also show promise for use in artificial neurons, because of the inherent stochasticity in their resistance changes.[13] This inherent stochasticity of memristive neurons can be leveraged for better signal representation,[14,15] or more efficient probabilistic computing than would be possible with deterministic neurons.[16] Nonetheless, applying memristive elements in artificial neurons is more complex and has been much less explored compared to their application in synapses.

Here, we demonstrate a simple memristive neuron based on a halide perovskite memristive element. Metal halide perovskites are semiconducting compounds that efficiently conduct both electronic and ionic charge carriers.[17] The efficient ion conduction in halide perovskites readily induces hysteresis, which was previously exploited to make energy-efficient artificial synapses.[18–20] While various halide perovskite artificial synapses have been reported, only one halide perovskite neuron has been experimentally demonstrated before.[21] However, this previous implementation used off-chip circuitry to implement signal integration and neuron-like spiking, making scaling difficult. We connect a microscale volatile halide perovskite memristive device in series with a capacitor. The series capacitor applies a reverse bias on the memristive element after spiking of the neuron, which aids in resetting the memristive element after each spike. This makes our neuron design more robust against non-reversible resistance changes of the memristive component than designs with a series resistor,[22,23] or capacitor connected in parallel.[16,24,25] Because our design consists of only two components, the neuron is also more easily scalable than implementations that require more complex circuitry besides the memristive element.[14,26,27] Moreover, the efficient ion conduction of halide perovskites allows an operating voltage of hundreds of millivolts, lower than in previous memristive neurons which is favorable for low energy consumptions. We fabricate our crosspoint neurons with a previously developed procedure that prevents degradation of the halide perovskite layer during lithography.[19] Our neuron is integrated fully on-chip without the need for external circuitry to emulate neuron functionality. In that way, the device architecture of our halide perovskite memristive device lends itself to further downscaling and the neuron could be easily integrated with halide perovskite artificial synapses that we have demonstrated before to form artificial neural networks with ultralow-energy consumption.[19]

# Experimental

## Fabrication of the on-chip artificial neuron

Heavily p-doped Si wafers (1-5 Ω · cm resistivity) were purchased from Siegert Wafer. $PbI_2$ (99.99%) was purchased from TCI. Methylammonium iodide (MAI) was purchased from Solaronix. Anhydrous DMF and chlorobenzene were purchased from Sigma-Aldrich. 950 PMMA A8 was purchased from Kayaku Advanced Materials. All materials were used without further purification.

Devices were fabricated using a similar procedure as described before.[19] The artificial neurons were fabricated on heavily p-doped Si wafers with a 100 nm thermal oxide layer. Gold bottom electrodes were patterned on the wafer with a lift-off procedure using MA-N1410 photoresist. UV exposure with a Süss MA6/BA6 mask aligner was followed by development in MA-D533/s. A 5 nm Cr adhesion layer and an 80 nm Au electrode layer were deposited on the patterned resist by e-beam physical vapor deposition. Lift-off was then performed by soaking in acetone for one hour. A 60 nm $SiO_2$ layer was deposited from a $O_2$ and $SiH_4$ gas mixture using inductively-coupled plasma-enhanced chemical vapor deposition (ICPCVD) in an Oxford PlasmaPro100 ICPCVD system. Silver top contacts were patterned using the same procedure as for the bottom electrodes. After patterning of the top electrodes, the $SiO_2$ layer was etched in an Oxford Plasmalab 80 Plus system with an Ar and $CHF_3$ gas mixture, using the top electrodes as a hardmask. Inside a nitrogen-filled glovebox (< 0.5 ppm $O_2$ and water), a stoichiometric mixture of $PbI_2$ and MAI was dissolved in DMF to obtain a 40 wt% $MAPbI_3$ precursor. The precursor was spin coated over the electrodes at 4000 rpm for 30 seconds in the same glovebox. Chlorobenzene was added as an antisolvent after 3 seconds of spinning. Directly after spin coating the samples were annealed at 100 °C for 10 minutes. The 950 PMMA A8 solution was spin coated on top of the halide perovskite at 3000 rpm for 45 seconds, followed by a 5 minute bake at 100 °C.

### Electrical characterization

I-V curves between -0.5 and 0.5 V and the retention time of the low resistance state were measured with a Keysight B2902A Precision Source/Measure Unit.

Artificial neuron measurements were performed by applying voltage pulses between the heavily p-doped Si substrate and the silver top-electrode with a Rigol DG1062Z arbitrary waveform generator, while measuring the voltage between the gold bottom electrode and the Si substrate with a PicoScope 6402C oscilloscope. The data was smoothed using a moving average with a 5 point subset, corresponding to a 20 μs time window. Afterward, 50 Hz noise from the AC power supply was removed using a fit to a sine wave with a 50 Hz frequency. Raw versions of the figures in the main text are given in Figure S11 and show that the measured signal is not affected significantly by the noise removal.

## Results and Discussion

Artificial neurons can be fabricated from a resistive switch that shows rapid, highly volatile switching connected in series with a capacitor.[28] Thereby, successive voltage pulses eventually switch the memristive element to the low resistance state, charging the capacitor (firing). Then, the charged capacitor reverse-biases the memristive element, switching it off again. We use a resistive switch that comprises of methylammonium lead triiodide (MAPbI$_3$) as the active layer, and a gold and silver contact as the bottom and top contact respectively (Figure 1a and Methods section). The 2.5 µm wide contacts are arranged in an overlapping back-contact geometry, where the two contacts are orthogonally placed on top of each other with an insulating spacer layer of SiO$_2$ in between. All lithographic processing steps are therefore performed before the perovskite deposition. The compact, dense structure lends itself to downscaling.[19] This resistive switch shows a unipolar behavior with a clear threshold voltage of about 0.3 V, where the resistance rapidly changes by four orders of magnitude from approximately 1 GΩ to 100 kΩ (Figure 1b). This resistance change is maintained for a short period only after switching off the voltage pulse, about 125 ms in the case of Figure 1c, a requirement for the fabrication of an artificial neuron. A histogram of retention times based on 40 measurements is given in Figure S2. In no case is the retention time more than 500 ms.

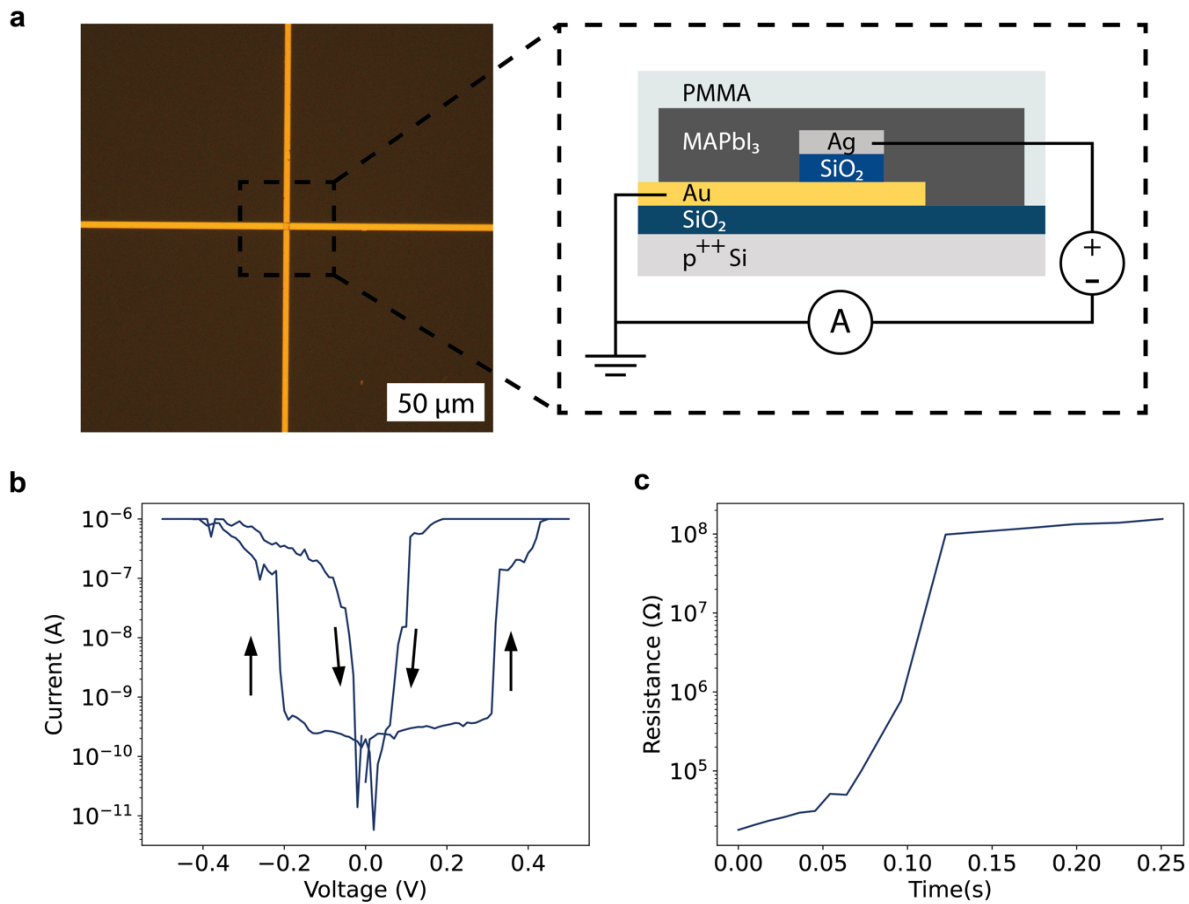

*Figure 1. A volatile halide perovskite resistive switch. **(a)** Optical microscopy image of the cross-point formed by the gold and silver electrodes before deposition of the halide perovskite layer, with a schematic image of the full resistive switching device. A gold bottom electrode and silver top electrode sandwich an $SiO_2$ insulating layer. Halide perovskite is spin-coated over the electrodes and forms the active layer of the device. **(b)** I-V curve of the device, measured between -0.5 and 0.5 V. The measured current increases by approximately 4 orders of magnitude at 0.3 V. The device returns to the initial high-resistive state as soon as the voltage is reduced to 0 V again and shows symmetric resistive switching properties in the negative poling direction. **(c)** Retention time measurement of the resistive switch. The resistance increases to that of the device in the high resistance state after approximately 125 ms. The full measurement is given in Figure S1.*

The resistance changes of the resistive switch are stochastic in nature, as is apparent from the histograms of the time to switch after applying the voltage pulse in Figure S3a, b, and c and their corresponding fit with a Poisson distribution. Such a Poisson distribution for the switching time is expected for resistive switches that change their resistance due to stochastic formation and destruction of conductive filaments.[13] We note that resistance change can also occur for the same device but without the $MAPbI_3$ layer, as illustrated by Figure S4. The switching then happens at about 10× higher voltages. It has previously been shown that silver filaments can form in $SiO_2$ layers,[29] and the resistance changes therefore likely occur due to filament formation through the $SiO_2$ spacer between the Ag and Au electrodes. Thus, the role of the halide perovskite layer in the final device is to strongly facilitate the formation of these Ag filaments, enabling lower voltage operation and thereby reducing the energy consumption of the device.

To turn this resistive switch into an artificial neuron, it needs to be connected to a capacitor. We implement this on-chip by connecting the resistive switch in series with a 300 pF capacitor that is formed by the Au bottom contact, the thermal $SiO_2$ layer and the highly-doped Si substrate, as shown in Figure 2a. With such a connection, the operation of the neuron follows three key steps, depicted in Figure 2b. In the first step, stimulation, the input voltage pulse experiences a resistive switch with high resistance. Therefore, every voltage pulse deposits only a small amount of charge on the capacitor, insufficient to build up significant voltage. After several pulses, the resistance of the resistive switch will promptly change to the low resistive state. At that point, the second step (firing) is initiated. The capacitor is quickly charged and the charge on the capacitor sets up a voltage that opposes the input voltage. The third step (resetting) is initiated when the applied voltage is removed. The capacitor discharges through the resistive switch, causing the resistive switch to return to the high resistive state, and the cycle can restart. Figure 2c shows the experimental realization of the spiking of the artificial neuron. A 33 Hz, 750 mV pulse train is applied to the device and the voltage across the capacitor is measured. We observe firing pulses on the capacitor after one to three applied pulses. Fitting of the charging and discharging of the capacitor in Figure S5a and b reveals that the resistance of the resistive switch is reduced to 1 to 4 MΩ during most firing steps. The resistance obtained from the fit is higher than the 100 kΩ obtained in the voltage sweep in Figure 1b, indicating that the device has not fully switched to the low resistance state. The voltage drop over the resistive switch is gradually reduced as the filament is forming and the capacitor is charged, leading to only partial formation of the filament. This partial formation of the filament further aids the volatility and energy efficiency of the device. During discharging of the capacitor in the resetting step, a resistance of approximately 10 MΩ is extracted, which corresponds to the input impedance of the oscilloscope. Assuming that the resistive switch is brought back to its 1 GΩ high resistance state during the resetting step, the oscilloscope provides a lower resistance discharge path for the capacitor, which is a limitation of our current measurement setup (see Figure S5c).

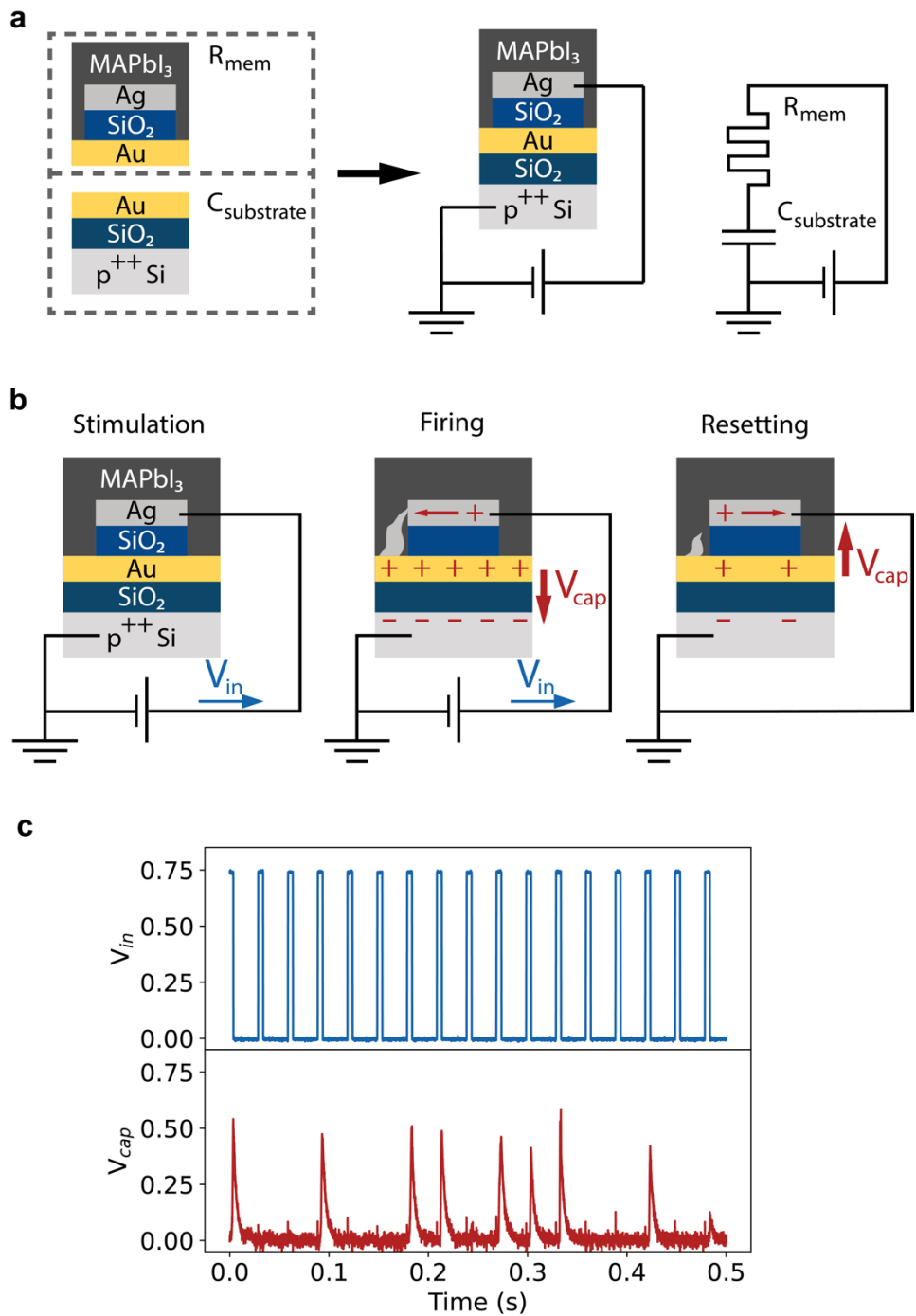

*Figure 2. Operation of the artificial spiking neuron. **(a)** The neuron is constructed by connecting the memristive part of the device, consisting of the gold bottom electrode, silver top electrode and the MAPbI$_3$ layer, with the capacitor formed by the gold electrode and contact pad, the 100 nm thermal SiO$_2$ layer and the highly doped Si substrate in series. **(b)** Schematic representation of the three stages of the operation of the neuron. Upon application of a voltage, the device first undergoes a "stimulation" phase, where there is no significant voltage build-up on the capacitor due to the high resistance of the memristive part of the device. After enough voltage has been applied to the device, the memristive device switches to the low-resistance state and the capacitor is rapidly charged, causing a voltage buildup on the capacitor, i.e. "firing" of the neuron. When the applied voltage is removed, the capacitor discharges. This reverse-biases the resistive switch, aiding the disruption of the conductive filament, called the "resetting" process. **(c)** A pulsed measurement of the artificial neuron. A pulse train of 5 ms, 0.75 V pulses are applied with a 33 Hz frequency, resulting in firing spikes on the capacitor.*

The capacitive discharge fit immediately corresponds to the oscilloscope impedance (Figure S5a), from which we conclude that the resistive switch is reset as soon as the bias is removed, at least on the timescale of the measurement. No firing pulses were measured if the halide perovskite layer was omitted, as shown in Figure S6. The resistance changes that underlie the spiking behavior of the neuron, therefore, occur through the halide perovskite layer at these low applied voltages. Figure S7 shows that the stochastic spiking of the neuron was reproducible over multiple measurements.

The firing pattern of the neuron is stochastic in nature, which is expected from the underlying stochastic switching mechanism of the resistive switch. Similar to the resistive switch itself, Figure S8a shows that the time under bias before spiking of the neuron follows a Poisson distribution, with a mean of 6.9 ms for the 0.75 V pulses. This stochastic switching is also observed in biological neurons and can have advantages compared to purely deterministic neurons.

To demonstrate this advantage we use the experimentally obtained mean switching time and resistances to model the behavior of the stochastic neuron. We compared the simulated stochastic neuron to a hypothetical deterministic neuron with a deterministic threshold of the same time constant (6.9 ms) to determine the ability of stochastic and deterministic neurons to represent the input voltage pulse train. Modeling of the neuron is discussed in more detail in Supplementary Note 1.

Figure 3a shows the simulated spiking behavior of a stochastic and a deterministic neuron. The spiking of the simulated stochastic neuron is similar to that in the measurement shown in Figure 2c. The simulated deterministic neuron, on the other hand, spikes at regular intervals.

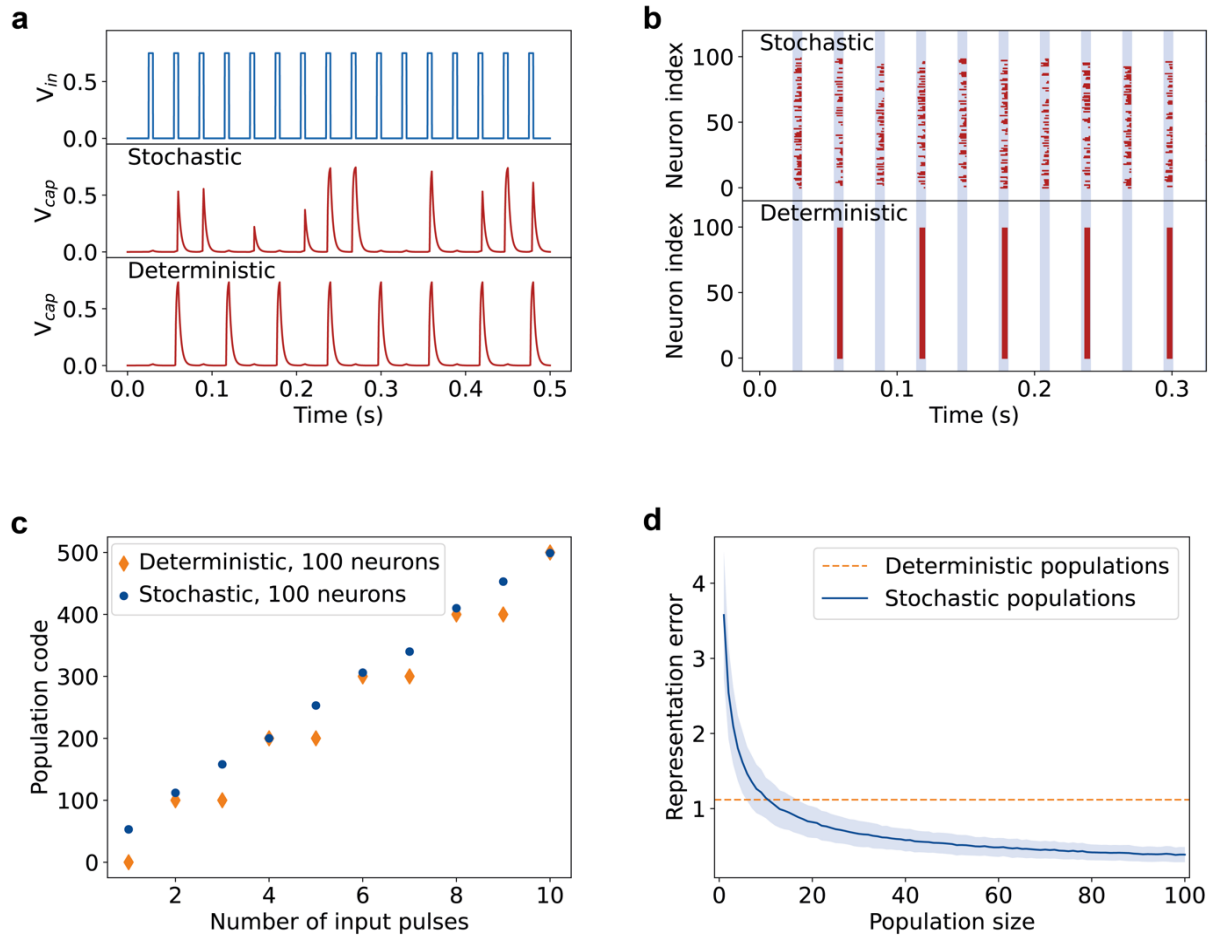

*Figure 3. Simulations comparing the stochastic spiking of the neuron with a hypothetical deterministic version of the neuron. **(a)** Comparison of a simulated stochastic and deterministic spiking neuron, with the same input as in Figure 2c. Similar spiking behavior is obtained for the simulated and experimentally measured stochastic neurons. The deterministic neuron always spikes after a cumulative 6.9 ms of bias has been applied. **(b)** Simulated spiking behavior of populations of 100 stochastic and deterministic neurons. Ten voltage pulses are applied in the simulation with the same pulse duration, length, and magnitude as **(a)**. Blue-shaded regions indicate the application of the 750 mV pulses, while the red marks indicate spiking by the neuron. While the deterministic neurons all spike at the same time, spiking by the stochastic neurons is distributed more evenly throughout the applied pulses. **(c)** The population codes obtained for each applied pulse in **(b)**. We define the population code as the cumulative amount of spikes output by the population. For the deterministic population, the population code increases with each even number of applied pulses, while the stochastic population shows a more gradual increase with each applied pulse. **(d)** The representation error of deterministic and stochastic populations as a function of the population size, averaged over 1000 simulations. Deterministic populations have the same representation error regardless of their size. The representation error of the stochastic neurons decreases as the population size increases. The representation error of the stochastic populations is lower for population sizes of 11 or more neurons. The blue shaded region indicates one standard deviation.*

To achieve more biologically plausible, robust, and accurate spiking neural networks, neurons are typically implemented in populations.[14,15] In these networks, input signals are fed into the neurons in the populations and their collective output is collected as a population code. Figure 3b shows a simulation of populations of 100 stochastic or deterministic neurons. While the spikes of the stochastic neurons are distributed over all input voltage pulses, the deterministic neurons spike uniformly roughly each second input pulse.

From the simulations of the stochastic and deterministic neuron populations, we calculate the population code as the cumulative amount of spikes output by the total population after each successive input pulse, Figure 3c. The population code for the deterministic populations increases stepwise, showing that the stochastic neurons can better distinguish different numbers of applied pulses, *i.e.,* they can better encode or represent the input. This process by which stochastic neurons can pick up on sub-threshold signals is called "stochastic resonance". Biological neurons, which are also stochastic, rely on stochastic resonance to detect otherwise sub-threshold signals.[30]

To study the effect of population size on the reliability of signal detection, we simulated population codes for populations of 1 and up to 100 neurons and computed a signal representation error for each population size, see Figure 3d. Supplementary Note 1 explains how the representation error was determined. This representation error measures how well the population can encode and distinguish between different inputs. The representation error is initially larger for small populations of stochastic neurons compared to deterministic ones. However, the error rapidly decreases as the population size increases and drops below that of the deterministic neurons for relatively small population sizes of 11 or more stochastic neurons. These results are in line with previous work where the same benefit was found for stochasticity in artificial neuron populations.[14,15]

Experimentally, the neurons are stochastic, but the stochasticity is tunable. The spiking behavior of the neuron can be tuned by changing the parameters of the input voltage pulses. As shown in Figure 4a, the neuron outputs spikes with a higher probability for each input pulse if the frequency of the incoming pulses is increased. On the other hand, a lower input pulse frequency in Figure 4b leads to no spiking of the neuron, which is a clear demonstration of the leaky behavior of the neuron. Another demonstration of the leaky-integrate-and-fire behavior of the neuron is given in Figure S10. Increasing the pulse duration to 7.5 ms leads to firing with each applied voltage pulse, whereas 2 ms pulses applied with the same frequency do not lead to spiking of the neuron.

Changing the voltage also provides a way to change the firing pattern of the neuron. When the neuron is integrated in full networks, this would be equivalent to connecting the neuron through synapses with a low connection strength, *i.e.* a high resistance. The measurement in Figure 4c illustrates that a lower voltage drop over the neuron due to a resistive artificial synapse leads to no spiking of the neuron. Our spiking neuron therefore shows the leaky-integrate-and-fire behavior and synaptic strength-dependent spiking properties required for constructing neuromorphic hardware with the synapse.

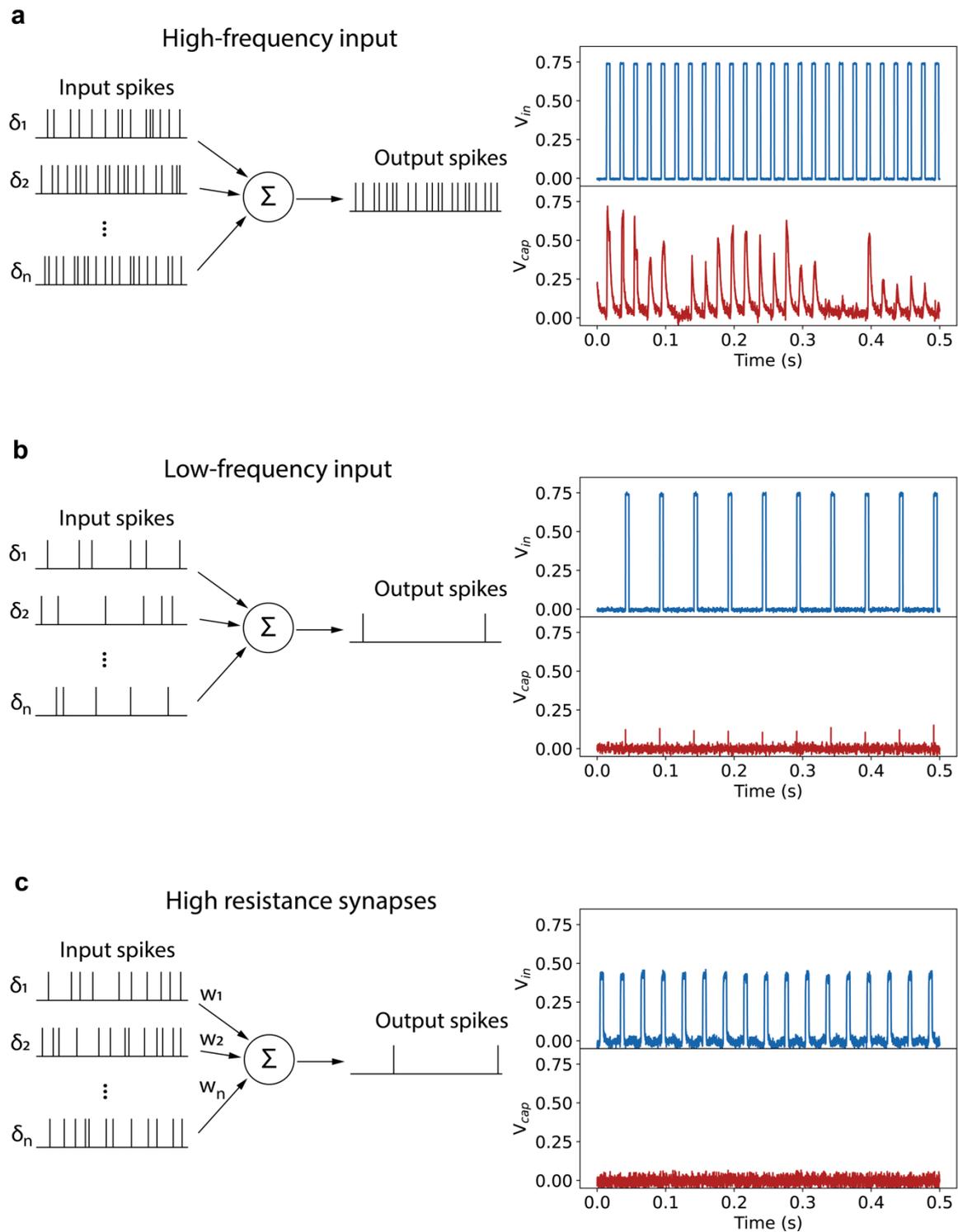

*Figure 4. Tunability of the firing of the neuron. **(a)** Increasing the frequency of the incoming voltage pulses to 50 Hz leads to a higher firing probability with each input pulse. **(b)** At a lower frequency of incoming voltage pulses of 20 Hz the neuron does not fire. **(c)** A lower input voltage of 400 mV, corresponding to connection of the neuron through high resistance synapses, leads to no firing of the neuron.*

The energy consumption of the firing pulses can be calculated by $E = \frac{1}{2} \cdot C \cdot V^2$, with $C$ the capacitance of the on-chip capacitor and $V$ the voltage of the firing pulse, which yields an energy consumption per firing pulse between 20 to 60 pJ. This is already lower than the energy consumed by a biological neuron (on the order of 100 pJ),[31] and artificial neurons that have been implemented in hardware spiking neural networks before,[32] even in this early adaptation. More energy-efficient silicon artificial neurons that were demonstrated before have not yet been implemented in full networks.[33] In addition, neurons based on electronic circuits of traditional transistors and capacitors require a large number of these components,[8,33] making the circuits bulky and therefore limiting the maximum density that can be reached on the final chip. In contrast, our design consists of only two components and could therefore be incorporated in higher densities more easily. Moreover, there is no detectable voltage build-up on the capacitor during the stimulation step before firing, meaning that the energy consumption per spike can be reduced by reducing the capacitance of the capacitor without negatively influencing the functioning of the neuron. We discuss further scaling effects in Supplementary note S2 in the Supplementary Information.

Biological neurons are sensitive to input signals of similar frequencies that we use in this work.[34] Although these frequencies are significantly lower than that of conventional computers, the different way that information is processed in neuromorphic networks still allows for efficient computation. In fact, neuromorphic networks require synapses and neurons that have time constants that are well-matched to their input for efficient computation. Thus, interfacing with the natural world, *e.g.* for learning from visual input, requires operating frequencies similar to those we use here.[7,35] These time constants can be difficult to achieve with CMOS-based neuromorphic hardware.[36] Our neuron therefore provides a convenient alternative that is natively capable of operating at these frequencies. The ability to incorporate these neurons and the corresponding artificial synapses on flexible substrates could allow for novel application areas, including soft robots or even in combination with biological tissue. In addition, ion conductivity and corresponding resistance changes of halide perovskites can be tuned by light stimulation.[37] Perovskite neurons could therefore also open up new possibilities of hybrid electronic-photonic neuromorphic hardware, such as low-power smart sensors.

# Conclusion

In conclusion, we have demonstrated the first fully on-chip halide perovskite artificial neuron. The neuron consists of only two components, which lends itself well to high-density integration, and shows clear leaky-integrate-and-fire behavior, important for integration in neuromorphic hardware. The spiking of the neuron is stochastic, similar to biological neurons, yet with a lower energy consumption per spike between 20 to 60 pJ. The stochastic spiking of the neuron is beneficial for detecting sub-threshold input, similar to biological neurons. The energy consumption of the neuron could be further reduced by lowering the capacitance of the capacitor. The similarity in device architecture of this artificial neuron to the downscaled artificial synapses of $MAPbI_3$ that we have shown before,[19] allows easy implementation of energy-efficient all-halide perovskite neuromorphic hardware.

# Conflicts of interest

There are no conflicts to declare.

# Acknowledgements

The work of J.J.B. and B.E. received funding from the European Research Council (ERC) under the European Union's Horizon 2020 research and innovation programme under grant agreement No. 947221. The work is part of the Dutch Research Council NWO and was performed at the research institute AMOLF. The authors thank Marc Duursma, Bob Drent, Igor Hoogsteder and Laura Juškėnaitė for continuous technical support.

Supplementary Information for

# Integrated Artificial Neurons from Metal Halide Perovskites

Jeroen J. de Boer[1], Bruno Ehrler[1,*]

[1]*Center for Nanophotonics, AMOLF, 1098 XG, Amsterdam, the Netherlands*
[*]*Correspondence: b.ehrler@amolf.nl*

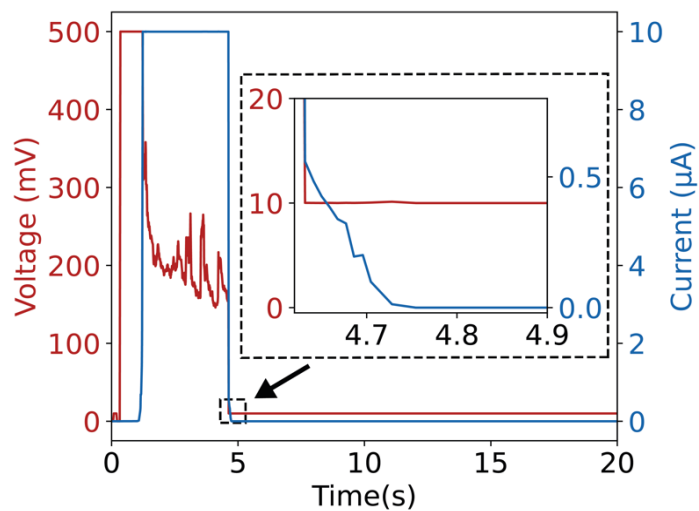

*Figure S1. Retention time measurement of the resistive switch. A 500 mV pulse is applied to the device to bring the device to the low resistance state. After approximately one second, the resistance of the device rapidly drops. The measurement setup then reduces the applied voltage to ensure that the set compliance current of 10 µA is not exceeded. After 4.5 seconds, the voltage is reduced to 10 mV to measure the evolution of the resistance over time. The inset shows a zoom-in on the region in the dotted rectangle, corresponding to the first hundreds of milliseconds after the potential is reduced to the 10 mV read-out voltage.*

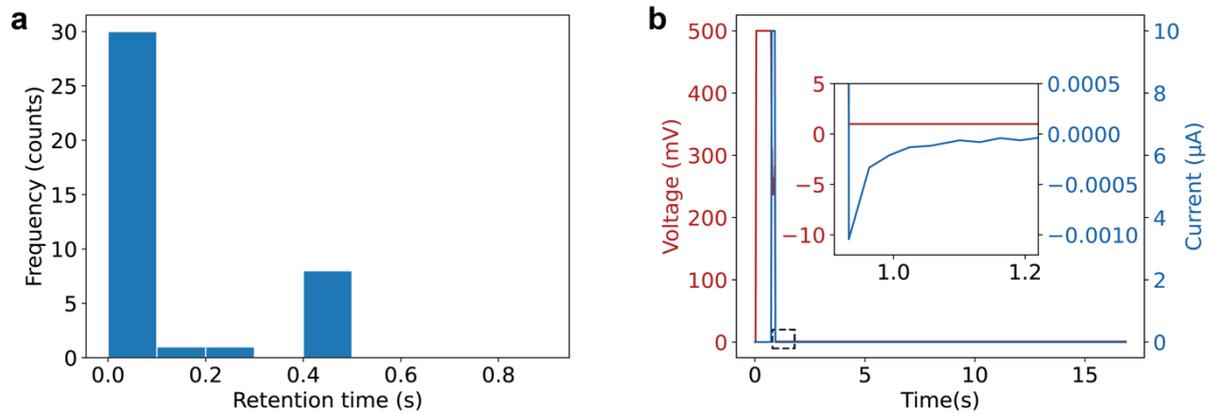

*Figure S2. Analysis of retention times of the resistive switch. **(a)** Histogram of 40 retention time measurements. **(b)** Example of a measurement where the retention time was less than the integration time of the measurement setup, so only a discharge of the parasitic capacitance is measured. Although the exact retention time of these measurements could not be determined, they were added to the first bin of the histogram to still give an accurate representation.*

Figure S2a gives a histogram of retention times based on 40 measurements. The device is set to the high-conductive state by applying 500 mV, with the compliance current set to 10 µA. The retention time was measured by applying 1 mV of constant bias. In several cases, the device reset to the low-conductance state within the integration time of the measurement setup (about 25 ms). An example of such a measurement is given in Figure S2b. These measurements were binned in the first bin (between 0 and 0.05 s) of the histogram in Figure S2a. The retention time was under 500 ms for all cases measured.

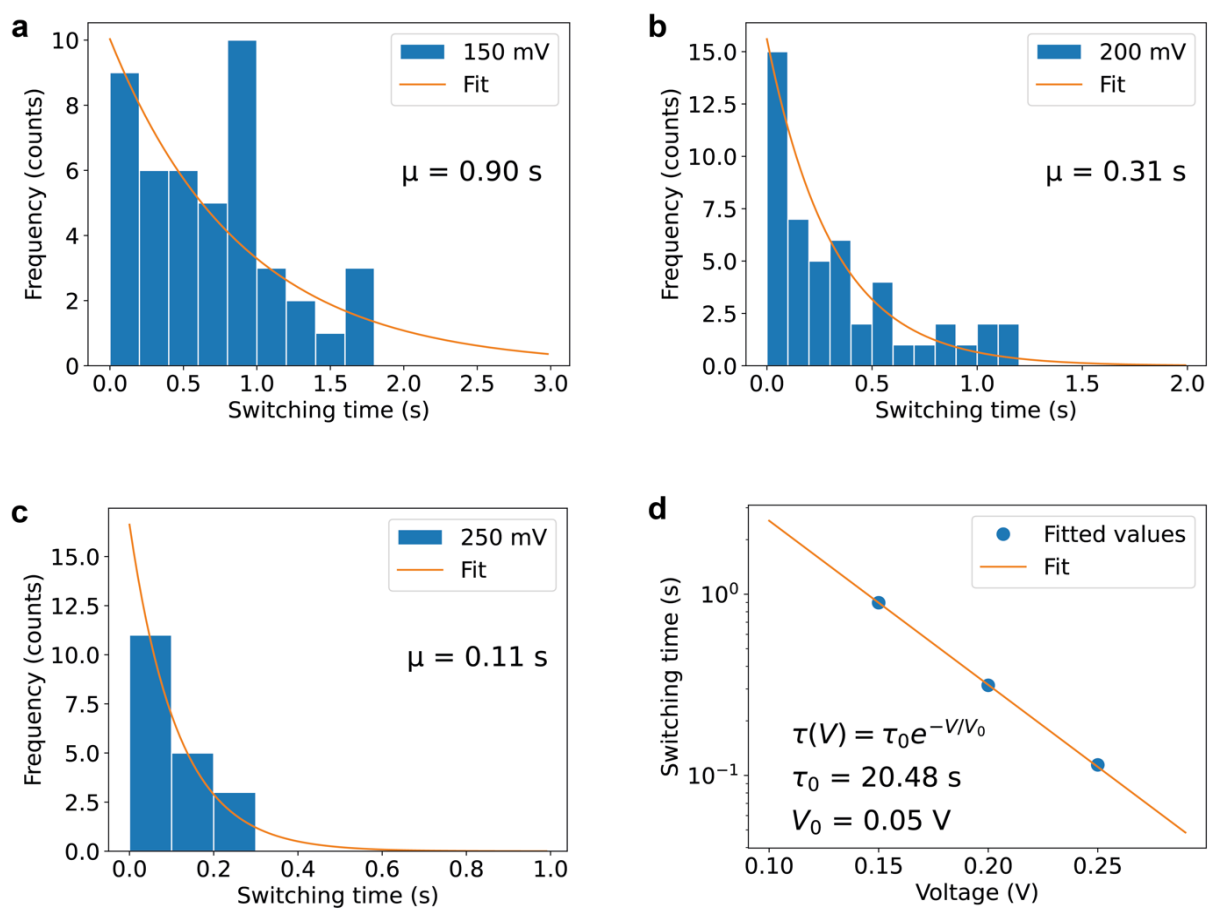

*Figure S3. Histogram of the switching time for a switching event of the halide perovskite memristive device under an applied voltage of 150 **(a)**, 200 **(b)**, and 250 mV **(c)**, with a fit based on a Poisson distribution. The means obtained from the fits are given in their respective figures. **(d)** A fit of the means obtained in the previous subfigures to the listed exponential function to extract the fitting parameters.*

The probability of a resistance change of the memristive device upon the application of a voltage follows a Poisson distribution, as is evident from Figure S3a, b, and c. The formation of conductive filaments in memristive devices requires hopping of ions by a thermally activated process, which introduces this stochasticity. Random fluctuations are not averaged out according to the law of large numbers due to the small amount of ions needed to form the nanoscale filament. Previous work has described this Poisson behavior extensively and showed that the mean switching time depends exponentially on the applied bias according to $\tau(V) = \tau_0 e^{-V/V_0}$, where $\tau$ is the mean switching time, and $\tau_0$ and $V_0$ are fitting parameters.[1] Figure S3d shows the same trend for our device, indicating that the same process of stochastic conductive filament formation underlies the operation of our device. Previous research on devices with similar electrodes and a halide perovskite active layer has shown that these conductive filaments consist of iodide vacancies[2] or silver.[3]

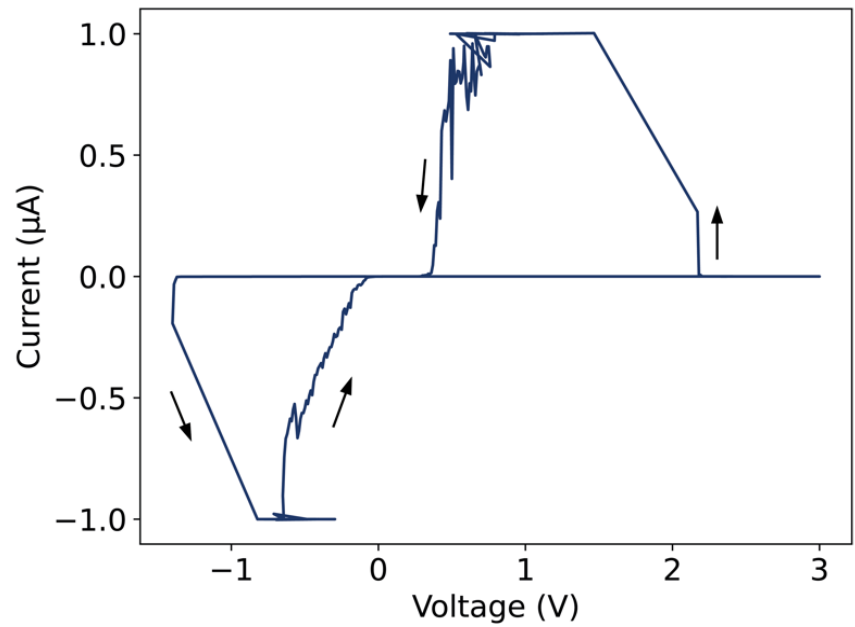

*Figure S4. Resistive switching of a device without the halide perovskite layer. Resistive switching also occurs through the SiO$_2$ spacer, albeit at higher voltages than for the device with a halide perovskite layer.*

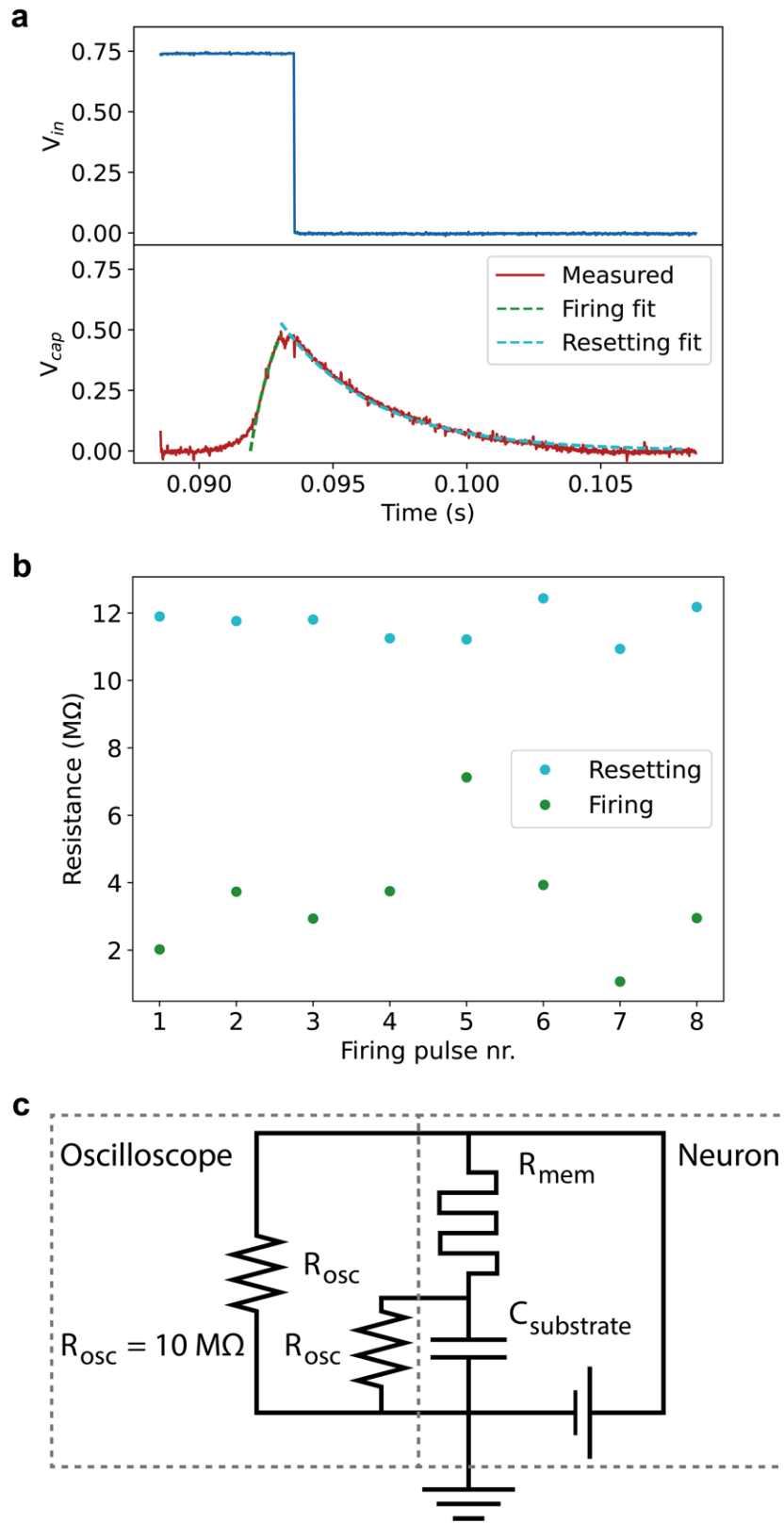

*Figure S5. Fit of the firing pulses in Figure 2c with charging and discharging of a capacitor. **(a)** Fits of the charging (firing) and discharging (resetting) of the second firing pulse. **(b)** Extracted resistances of the charging and discharging of each of the firing pulses in Figure 2c. Error bars representing one standard deviation of the obtained resistance from the fit are included, but are smaller than the dots of the markers in the figure. **(c)** The*

*circuit of the neuron with the 10 MΩ probes of the oscilloscope connected. The probes offer an alternative path for the capacitor to discharge.*

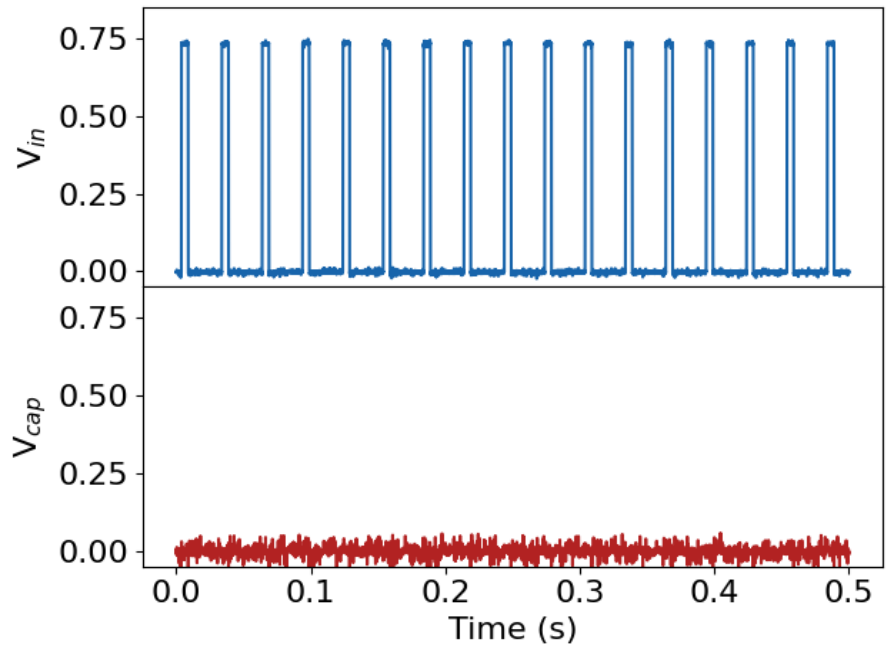

*Figure S6. Spiking neuron measurements repeated on a substrate without the halide perovskite layer. No spiking is measured using the same parameters as for the spiking neuron in the main text.*

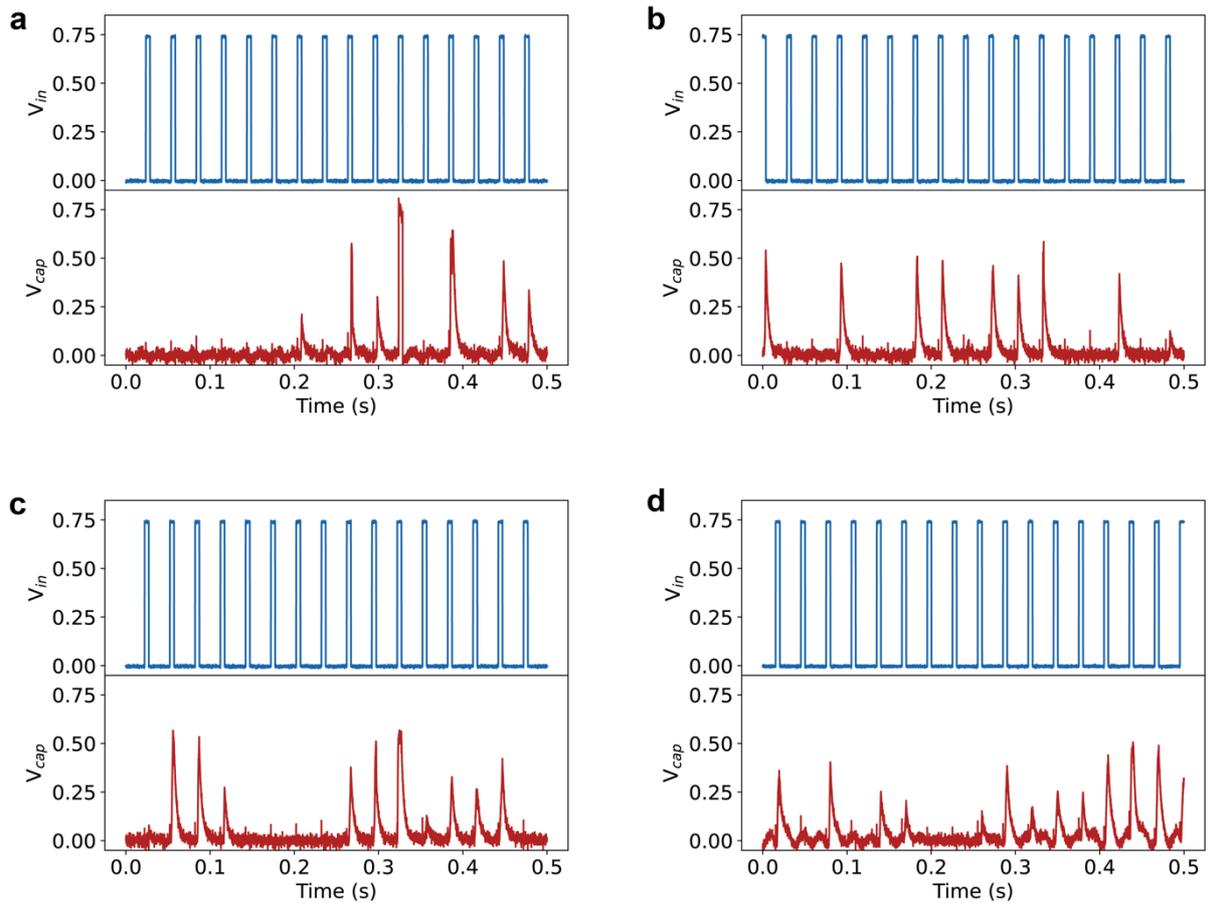

*Figure S7. Four different measurements of the spiking neuron. In the measurements in **(a)**, **(b)**, **(c)**, and **(d)** the same voltage profile, with 5 ms pulses of 750 mV, was applied to the neuron with a 33 Hz frequency. In all cases, this resulted in stochastic spiking by the neuron.*

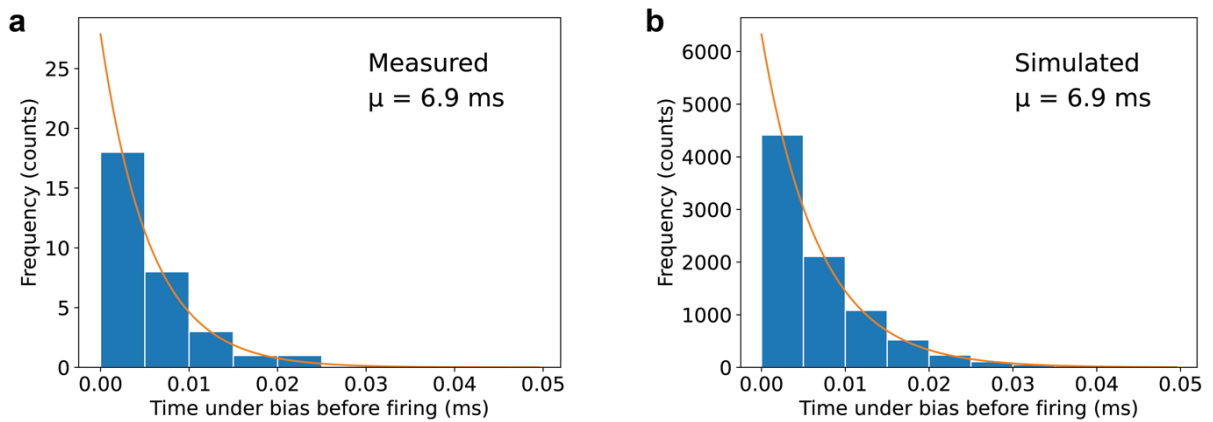

*Figure S8. **(a)** Histogram of the time under bias before firing of the neuron based on the measurements in Figure S7. Spiking by the neuron was defined as the moment when the capacitor voltage exceeds 200 mV. The mean of the distribution is 6.9 ms. **(b)** The same histogram of spiking by the neuron compiled from simulated data to validate the model. We obtain the same mean of 6.9 ms of applied bias before spiking by the neuron.*

# Supplementary Note 1. Modeling of stochastic and deterministic neurons

The simulated data in Figure 3 was obtained with experimentally determined resistances and capacitance. For the stochastic neurons, a switching time was drawn from a Poisson distribution with the experimentally determined mean of 6.9 ms from Figure S8, using a random number generator. Voltage pulses are applied in the simulation, which causes a voltage buildup on the capacitor with an RC constant of $300 \text{ pF} \times 1 \text{ G}\Omega = 0.3 \text{ s}$. The time under bias is then tracked until the switching time is reached. At this point, the RC constant decreases significantly to $300 \text{ pF} \times 3 \text{ M}\Omega = 0.9 \text{ ms}$ due to the resistance change of the resistive switch. After the voltage is removed, the capacitor discharges through a 10 MΩ resistor and a new switching time is drawn from the Poisson distribution. The same procedure is followed for deterministic neurons, but with a switching time always set to 6.9 ms. For validation of the model, we simulated the spiking of a neuron using the same input voltage profile as for the measurements in Figure S7, but over a longer period of 500 seconds, constituting 16,666 applied voltage pulses. A histogram of the time under bias before spiking by the neuron based on the simulation is given in Figure S8b. The mean obtained from the simulation of 6.9 ms agrees with the experimentally obtained mean of 6.9 ms.

To determine the representation error of the neuron populations in Figure 3d, the population code of the neuron populations, *i.e.,* the cumulative sum of spikes for each applied pulse, is compared to the ideal population code. For stochastic neurons, the chance of the neurons outputting a spike is $P(5 \, ms) = 1 - e^{-\frac{5 \, ms}{6.9 \, ms}} \approx 0.52$ for each 5 ms input voltage pulse. In the ideal case, the mean number of spikes per neuron in a population is, therefore, equal to 0.52 multiplied by the number of applied pulses. Figure S9a compares the ideal mean number of spikes with that of a simulation of stochastic neuron populations of different sizes. Because of the non-zero chance of spiking by the stochastic neuron for each applied pulse, the population can capture all applied voltage pulses. For larger population sizes, the mean of the pulses approaches that of the ideal case. The deterministic neurons always spike after 6.9 ms of bias, or every second pulse.

Because of the deterministic nature of the neuron, additional spikes are never output by deterministic populations for uneven numbers of applied pulses, as illustrated by Figure S9b.

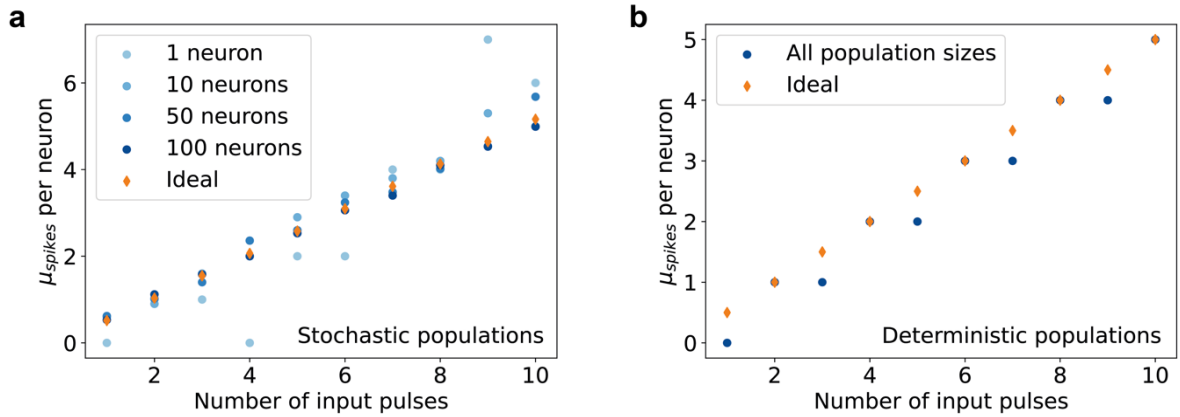

*Figure S9. The mean number of spikes per neuron of the stochastic and deterministic neuron populations, compared to their respective ideal mean number of spikes. **(a)** The mean number of spikes per number of stochastic populations of different sizes. Owing to the stochastic nature of the spiking, the average number of spikes converges to the ideal case for all applied pulses as the population size increases. **(b)** The mean number of spikes of deterministic populations. By definition, the mean does not change for different population sizes. The mean does not increase for uneven number pulses but matches perfectly with the ideal mean for even numbers of applied pulses.*

To obtain the representation error from the simulations, we take the Euclidian norm of the difference between the ideal and the simulated mean numbers of spikes for each population, $E = \sqrt{\sum_{i=1}^{10}(\mu_{i,simulated} - \mu_{i,ideal})^2}$, where $i$ refers to the applied pulse number and µ is the mean number of spikes per neuron. To obtain the results shown in Figure 3d, we repeated the simulations 1000 times for neuron populations of 1 and up to 100 neurons. The average representation error is shown in the figure.

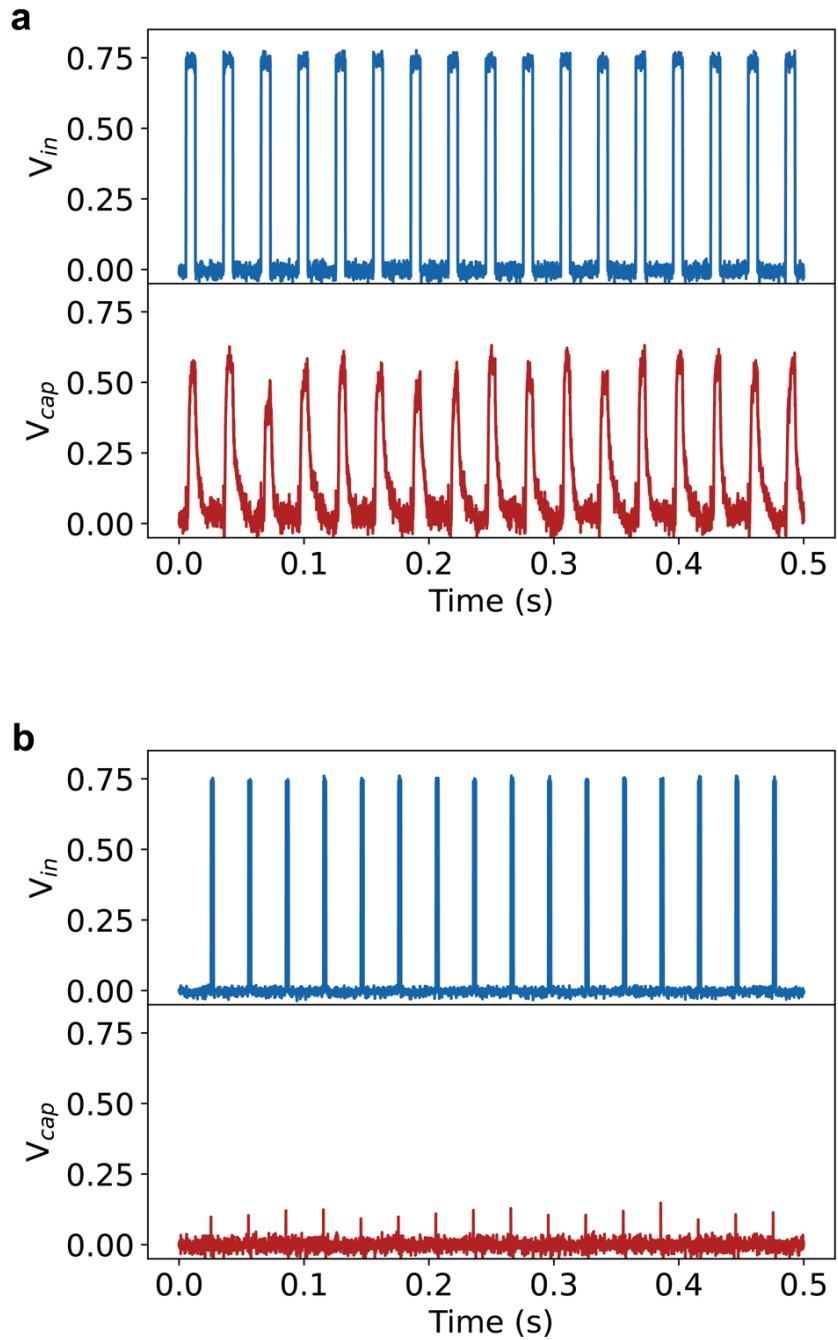

*Figure S10. Stimulation of the neuron with 750 mV pulses with pulse durations of 7.5 ms in **(a)**, which leads to firing with every applied pulse and with a duration of 2 ms in **(b)**, which leads to no firing of the neuron.*

# Supplementary Note 2. Scaling of the spiking neuron

There are two main limitations of further scaling of the neuron. First, the RC constant should remain high enough to prevent excessive charge buildup on the capacitor during the stimulation phase. To estimate scaling limitations, we set a limit of 100 mV of voltage buildup during stimulation of the neuron, while the resistive switch is still in the OFF-state. A voltage buildup of 100 mV is still easy to distinguish from a firing event where the voltage rises to several hundreds of millivolts, as in the measurements in Figure S7. Assuming the same input pulses of 750 mV and 5 ms in length as in Figure 2c, the RC constant should then be at least $RC = -\frac{t}{\ln\left(1-\frac{V(t)}{V_{supply}}\right)} = -\frac{5 \times 10^{-3} s}{\ln\left(1-\frac{100\ mV}{750\ mV}\right)} \approx 35\ ms$. Further downscaling of the resistive switch should linearly increase its low-conductance state resistance. Balancing this resistance with the series capacitance to satisfy the constraint on the RC constant allows for easy scaling of the neuron.

Even if the low-conductance state resistance is not reduced further as the resistive switch is scaled, due to a parasitic resistance in the circuit, for example, the capacitance can still be reduced by approximately an order of magnitude, to $C = \frac{.035\ s}{10^9 \Omega} = 35\ pF$, assuming an OFF-state resistance of 1 GΩ extracted from the I-V curve in Figure 1b. Assuming that the reduction in the capacitance means that the capacitor is now always fully charged with a firing event, this would reduce the energy consumption of the neuron to $E = \frac{1}{2} \times C \times V^2 = \frac{1}{2} \times 35 \times 10^{-12} F \times (0.75\ V)^2 \approx 9.8\ pJ$. Even in this upper limit, the energy consumption of the neuron would be close to that of the most energy-efficient silicon neurons.[4]

A second limitation is that the amount of charge on the capacitor should be large enough to aid the resetting of the resistive switch after firing. From Figure S5a we see that the device is in the high resistive state right after the voltage is turned off. Thus, the switching time is <1 ms. The neuron would work in a similar way with a switching time 2-3 orders of magnitude slower, which means that a capacitance 2-3 orders of magnitude smaller would suffice, in the hundreds of femtofarad regime. Common CMOS technology utilizes architectures that can fabricate capacitors on this scale with very small device footprint.[5]

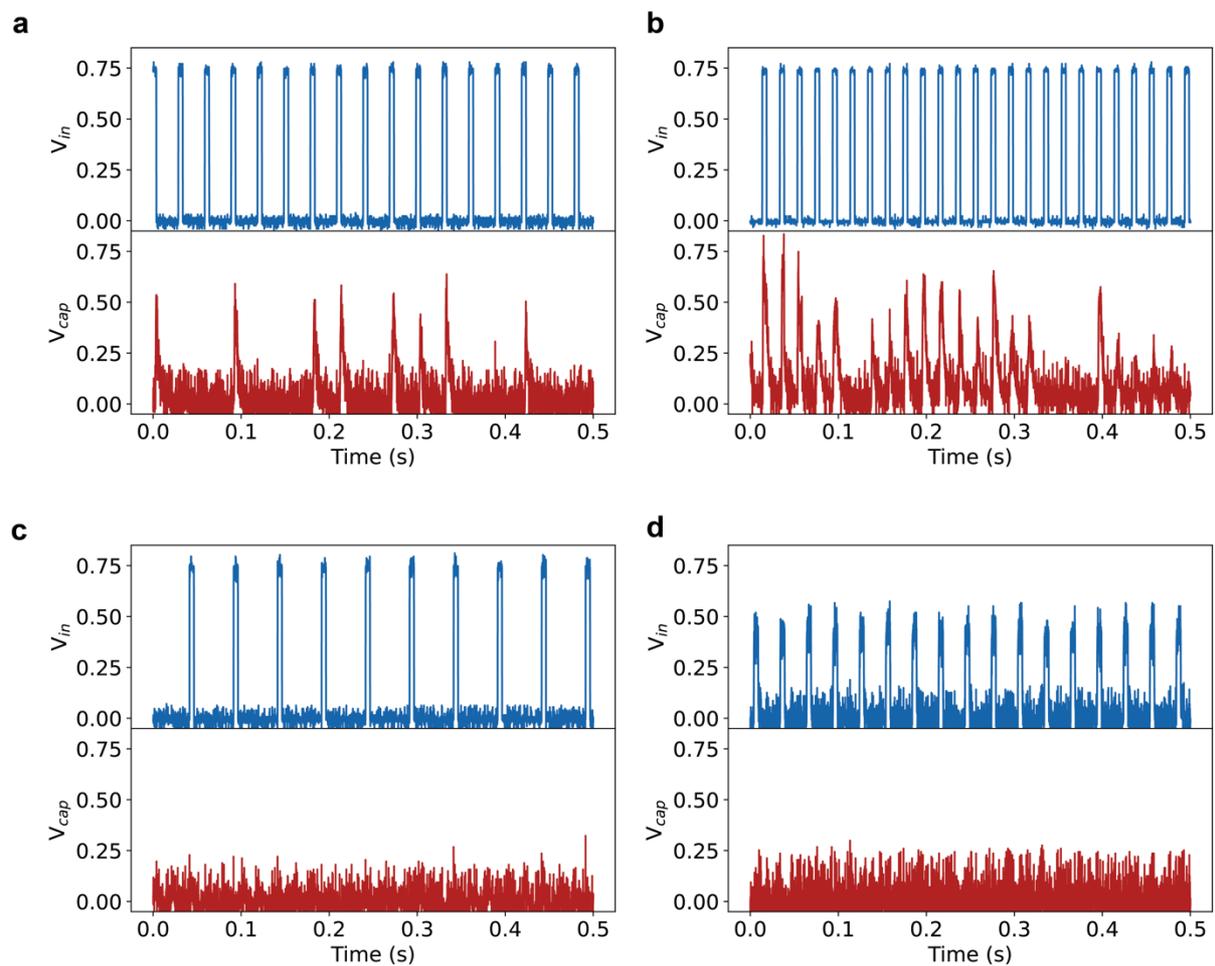

*Figure S11. **(a)** Figure 2c, **(b)** Figure 3a, **(c)** Figure 3b and **(d)** Figure 3c before smoothing and removal of the 50 Hz signal of the AC mains. Comparing these figures with the figures in the main text shows that the noise is removed without distorting the measured signal.*